# Optimizing Search Strategies: A Study of Two-Pointer Linear Search Implementation

## Nazma Akter Zinnia[1] and Eisuke Hanada[2]

1. Graduate School of Science and Engineering, Saga University, Saga 8408502, Japan; zinnia.cse7862@gmail.com
2. Faculty of Science and Engineering, Saga University, Saga 8408502, Japan; hanada@cc.saga-u.ac.jp

**Abstract:** A search algorithm is used to retrieve information stored in a group of data and is calculated in the search space of a problem domain, either with discrete or continuous values. The purpose of a search is to extract data from a large dataset, and there are two main types:
    1. Linear search and
    2. Non-linear search.
A linear search is a sequential process while a non-linear search doesn't follow any sequence. Linear and non-linear search algorithms are given below. The average run time of a linear search is O(n) and the best case complexity O(1). In a non-linear search the average run time complexity is O(logn) and best case complexity O(1).
With a binary search algorithm, the data needs to be sorted, which makes for extra time complexity. The two pointer technique provides better time complexity than a normal linear or binary search, and it works with sorted and in some cases unsorted data. A two pointer search is faster than a normal linear search. The time complexity of a two pointer search is the same as a normal linear search for the best case O(1) and average case O(n). Our proposed algorithm can be used with a large dataset where we have to find specific data and a matching item: the targeted sum and targeted element.

## 1. Introduction

A linear search is a way to find something in a list. Unlike a binary search, it doesn't need the list to be in any particular order. It works like this: you start at the beginning or end of the list and check each item until you find what you're looking for. If the array traverse is complete but there are no matches to the given condition, there are no items fitting the condition in that array. Linear search time complexity is O(n).

Because a linear search goes from left or right and iterates over the array items one by one, it can take much time in worst-case scenarios. In our thesis, we show how to optimize this time complexity, which can greatly reduce the processing time, even in a worst-case scenario. In the first part we show the basic principles of a linear search and its uses. In the second part we show how to improve linear searches by the use of the two pointer search technique.

## 2. Materials And Methods

### 2.1 Hardware Requirements

To be used efficiently, all computer software needs certain hardware components or other software resources to be present on a computer. These prerequisites are known as (Computer) system requirements and are guidelines as opposed to absolute rules. The system requirements for most software are stated as minimum and recommended. The following tables list the minimum and recommended hardware requirements for deploying two pointer search algorithms.

**Table 1: Minimum and Recommended Hardware for the Proposed System**

| Components | Minimum | Recommended |
|---|---|---|
| Processor | Intel Pentium | Core i3 CPU @ 1.60GHz |
| RAM | 4GB | 8GB |
| Storage | 500GB HDD | 1TB |

### 2.2 Software Requirements

Software Requirements deal with defining software resource requirements and prerequisites that need to be installed on a computer to provide optimal functioning of an application.

Our system was built for a Microsoft windows operating environment, and it is compatible with versions of windows that run client version 7 or later. The following Windows operating system editions can be used-
- Windows Operating System 8, version no. NT 6.2
- Windows Operating System 10, version no. NT 10.0
- Linux Operating System (Ubuntu), version no. 20.02

## 3. Existing Search Algorithms

Search Algorithms search for an element in a given array with either a linear or non-linear approach. For our non-linear example we will describe a binary search because the dataset used in our thesis contains only numeric values: there are no items such as graphs or trees.

## 3.1. Linear Search Algorithm

Here n is the size of the array, i is the index, and x is the searchable value or desired value. A is the array.

Step 1: Set i to 1
Step 2: if i > n, go to step 7
Step 3: if A[i] = x, go to step 6
Step 4: Set i to i + 1
Step 5: Go to Step 2
Step 6: Print Element x Found at index i and go to step 8
Step 7: Print element not found
Step 8: Exit

## 3.2. Binary Search Algorithm

For the binary search, 'a' is the given array, 'lower_bound' is the index of the first array element, 'upper_bound' is the index of the last array element, and 'val' is the value being searched for.

Step 1: set beg = lower_bound, end = upper_bound, pos = - 1
Step 2: repeat steps 3 and 4 while beg <=end
Step 3: set mid = (beg + end)/2
Step 4: if a[mid] = val
    set pos = mid
    print pos
    go to step 6
    else if a[mid] > val
    set end = mid - 1
    else
    set beg = mid + 1
    [end of if]
    [end of loop]
Step 5: if pos = -1
print "value is not present in the array"
[end of if]
Step 6: Exit

## 3.3. Comparison of Binary and Linear Searches

i. A binary search is faster than a linear search, but it has some limitation such as only working if the data is sorted. Additional complexity in needed to sort the data. To run an array over unsorted data with a binary search would take more time than a linear search because it needs to sort the data first.

ii. A linear search will be faster than a binary search because it searches one by one no matter if the data is sorted or not, but a binary search is faster than a linear search for sorted data.

## 4 Proposed System Design
### 4.1. Proposed Technique

The algorithm reads an array from the beginning, then n = array size and two pointer i = 0 and j = n-1 and one newarr = [] to find duplicate values.

After that, both pointers check in a loop until the 1st pointer is greater than the last pointer. If any pointer finds the targeted value, it then checks to see if there are any repeated values. If any, it will push the repeated value to new array[]. Otherwise, the algorithm will stop searching.

If neither pointer finds the target value, the first pointer will increase by one digit, i++, and the other pointer will be decreased by one digit, j--. The process of the algorithm is shown in the flowchart in Figure 1.

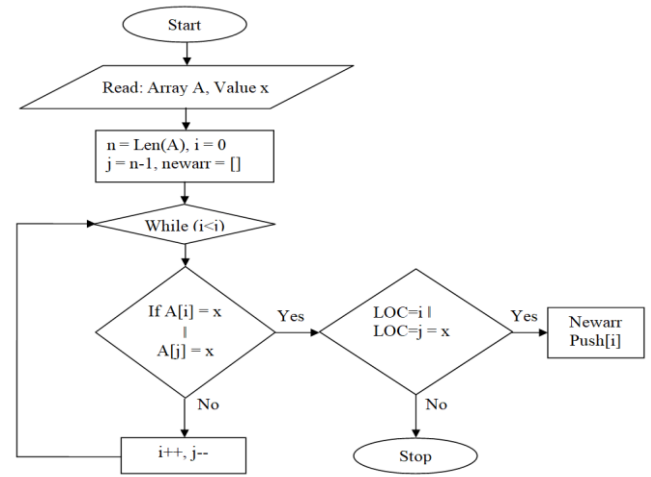

Figure 1: Proposed System Architecture

### 4.2. Proposed Algorithm

The proposed algorithm can be designed in two ways. First, we implement it to search for a unique or non- repeated value.
Here, n = array size
A = array
x= search value
i, j = pointer
newarr = [] array for storing a duplicate value

Step 1: Set i to 1 and j = n-1
Step 2: if i >j or j < I, go to step 7
Step 3: if A[i] = x or A[j] = x, go to step 6
Step 4: Set i to i + 1 and j to j-1;
Step 5: Go to Step 2
Step 6: Print Element x Found at index i and go to step 7
Step 7: Check the element in the full array if x is found, then push the element into newarr[]. If not found go to step 9.
Step 8: Print element not found
Step 9: Exit

For the second application we start with an array and a targeted value to be searched for in the array.

Step-1: Declare the two pointers: The first index i index or i pointer and the other index j index or j pointer.
Step-2: If the start index is greater than the end index or end index is lower than the start index, after the full array traverse we will not get the value we are searching for, so print not found at step 7.
Step-3: If the start index or end index matches the target value, it will print as step 6 then execute step 8: exit.
Step-4: Every iteration will increase by the value 1 and the value of j will decrease by 1.
Step-5: Running the process in a loop will move it to step 2 again and again.

**Complexity of the Proposed Algorithm:**
The complexity of the proposed algorithm has a given value.
**Best case:** In the best case the time complexity is $O(1)$
**Worst case:** In the worst-case scenario the complexity is $O(n)$
**Average case:** In an average case the complexity is $O(n)$

### 4.3. Comparison

A comparative analysis of the Linear, Binary, and Two Pointer techniques was done to evaluate the efficiency and performance of each algorithm across various scenarios to identify their respective strengths and weaknesses, thereby elucidating optimal use cases for each.

**Table 2: Comparison of Linear, Binary and Two Pointer Search Algorithms**

| Characteristics | Linear Search | Binary Search | Two Pointer Search |
|---|---|---|---|
| Definition | Starts searching from the element and searches until no items are found. | Finds the position of a searched for element by finding the middle element of an array. | Pointers search from each end. |
| Sorted Data | Not necessary | Must be sorted | Doesn't need sorted data. |
| Approach | Based on a sequential approach. | Based on a divide and conquer approach. | Based on a sequential approach. |
| Size | Preferable for a small data set. | Preferable for a large data set. | Fit for both small and large data sets. |
| Efficiency | Not efficient for large data sets. | Efficient for large data sets. | Efficient for large data sets. |
| Worst Case scenario | The worst-case scenario for finding an element is $O(n)$. | The worst-case scenario for finding an element is $O(\log_2 n)$. | The worst-case scenario for finding an element is $O(n)$. |
| Array Dimension | Works for 1D and 2D arrays. | Works for only a 1 dimensional array. | Works for 1D and 2D arrays. |
| Speed | Slow | Fast | Fast |
| Time complexity | $O(n)$. | $O(\log n)$. | $O(n)$. |
| Work procedure | Performs equality comparisons. | Performs ordering comparisons. | Performs less than and greater than comparisons. |
| Example | For an array containing 10 sorted data, a search for the 9th element takes 9 comparisons with a Linear search. | For an array containing 10 sorted data, a search for the 9th element takes 4 comparisons with a binary search. | For an array containing 10 sorted data, a search for the 9th element takes 2 comparisons with a two pointer search. |

### 5. Implementation

A case study on the impact of different devices for a fixed targeted value when using a linear search, binary search, or two pointer search of a one million item dataset is given below.

### 5.1. The Impact of Hardware and Software

Testing was done with the same, unique, unsorted dataset of one million integers. Specifications of the four devices used to test the algorithm. are given below.

**Table 3: Specifications of Hardware and Software used in the Search**

| Specification | Device 1 | Device 2 | Device 3 | Device 4 |
|---|---|---|---|---|
| RAM | 4GB | 4GB | 4GB | 8GB |
| Processor | Corei5 | Corei3 | Corei5 | Pentium Dual Core |
| OS | Windows 10 | Ubuntu 20.12 | MacOS Sierra | Windows 8 |
| Storage | 250GB SSD | 120GB SSD | 240GB SSD | 500GB HDD |

### 5.2. Dataset Preparation Script

First, we generated a file and named it dataset.txt. We ran a loop inside a loop and appended the file by inserting each item.

### 5.2.1. Sorted Data Preparation Script

The sorted data had integer numbers ranging from 1 to 10,000,000.

```
const fs = require("fs");
for (let i = 1; i <= 1000000; i++) {
 const content = i.toString() + "\n";
 try {
  fs.appendFileSync("dataset.txt", content);
 } catch (err) {
  console.error(err);
 }
}
```

### 5.2.2. Unsorted Data Preparation Script

Javascript was used to generate unsorted, random data. We first created a file and named it dataset.txt. We ran a loop inside a loop to generate random numbers ranging from 0-10,000,000, then appended the file by inserting every item into our dataset.txt file.

```
const fs = require("fs");
for (let i = 1; i <= 1000000; i++) {
 let x = Math.floor(Math.random() * 1000000 + 1);
 const content = x.toString() + "\n";
 try {
```

```
    fs.appendFileSync("dataset.txt", content);
  } catch (err) {
    console.error(err);
  }
}
```

**5.3. Result & Discussion**

**Result**

The hardware and software configurations significantly impacted the efficiency of the search algorithms, as shown in Table 4:

**Table 4: The Impact of Hardware and Software by Search Technique**

| OS | RAM | Data set | Targeted value | Linear Search | Binary Search | Two pointer linear Search |
|---|---|---|---|---|---|---|
| Windows 10 | 4GB | 1M | 31220 | 2 ms | 5 ms | 1 ms |
| Ubuntu 20.12 | 4GB | 1M | 31220 | 3 ms | 5 ms | 2 ms |
| MacOS Sierra | 4GB | 1M | 31220 | 2 ms | 6 ms | 2 ms |
| Windows 8 | 8GB | 1M | 31220 | 1 ms | 4 ms | 0 ms |

The tests on a variety of machines showed that our proposed two pointer algorithm is faster than the traditional search algorithms and that the speed depends on the processor and memory of the machine.

Our results also show that the new two pointer linear search algorithm works well for all input values and reduces complexity and that it works with both sorted and unsorted data. For a unique value, it gives the expected result in the first iteration, regardless of if it is from the left or right side. It can search the whole dataset for multiple values. It works with a variety of devices, whether they have the minimum required specifications or are more highly advanced systems.

**Discussion**

In this thesis we have proposed a two pointer search algorithm that proved to be much faster than algorithms for linear and binary searches and that has significantly improved time complexity.

We expected that our search algorithm would be faster than other methods, especially for big sets of data or sorted lists. Our testing showed that it did work faster than a linear search and was sometimes as fast as a binary search.

A two pointer search algorithm works by using two pointers, one at the start and the other at the end of a sorted list, that narrow down the search area until they find the targeted value. This method is good for sorted lists and is easier to use than other, more complicated algorithms. A binary search can be used with sorted lists but keeps dividing the list in half. Both methods are fast, but a binary search needs random access to the list elements. A linear search, however, goes through each item one by one, which is simpler but slower, especially for big lists.

Even though the two pointer search algorithm has advantages, it is not perfect. It works best with sorted lists, and sorting the list first takes time. While it is usually faster than a linear search, in some cases it is not faster than binary search, especially for very large lists. Also, if the list isn't sorted, the algorithm might not work well and becomes slower. It is important to consider these things when choosing which search method to use.

The two pointer search algorithm is good for tasks where the list is already sorted or can be sorted quickly. This makes it great for searching databases with previously sorted data, adding search features to algorithms like binary search trees, and looking for things in sorted lists. It can also be handy when we frequently change items in lists, keeping them sorted. By using the two pointer search algorithm, searching is faster and less computing power is necessary, especially with big or sorted lists.

**6. Conclusion and Future Work**
**6.1. Conclusion**

Searching means looking for something in a list of things. In a linear search, it is not necessary to have the list in any special order. Our results show that our new method of two pointer linear searching works well for all kinds of lists, which makes searching simpler. This new method works with lists that are in order or not. It can find what you're looking for in the list even if it is the first thing it checks, whether it is on the left side or the right side. It can search through the whole list for many things at once. It works on older devices as well as new ones. In this paper we introduced our modified version of a two pointer search technique. We were able to show how much faster this new method is compared to existing ones like linear and binary searches through the use of a variety of lists. We also created diagrams to explain how it works.

**6.2. Future Work**

Our two pointer linear search algorithm is very good at a wide variety of problems. There is no need to sort the data and it functions both forward and backward, which would make it very effective for use in the software development sector. We have worked with one of the search algorithms most widely used in software engineering, but we want to also be able to use it for real life applications.

Due to the limitation of device resources, we could not test more examples of our proposed algorithm, for example when used with graphs and trees, but plan to do this in the future. We plan to create software that can be use in real life based on this algorithm.

**Consent for Publication**
No Consent is required as our work is not a human study.

**Conflict of Interest**
The authors declare no conflict of interest.

## Acknowledgement

I would like to express my deep gratitude to the Almighty and my sincere appreciation to my thesis supervisor, Professor Eisuke Hanada, for his able guidance and supervision, which helped me to complete this thesis. I am extremely grateful and remain indebted to him for being a source of inspiration and for his constant support of the concept of the thesis. I would also like to thank those who have rendered me significant encouragement, vigorous support, and continuous and kind guidance. I thank them all for their invaluable guidance and other people who directly or indirectly assisted us in the successful and timely completion of this Two Pointers Linear Search Algorithm.

## References

[1] D. Knuth, "The Art Of Computer Programming," Sorting And Searching, vol. 3, Addison-Wesley, 1998.

[2] G. V. López, T. Gorin, and L. Lara, "Simulation Of Grover's Quantum Search Algorithm In An Ising-Nuclear-Spin-Chain Quantum Computer With First- And Second-Nearest-Neighbour Couplings," Journal Of Atomic, Molecular And Optical Physics, vol. 6, no. 4, pp. 123-135, 2008.

[3] A.H. Hunter and N.W. Pippenger, "Local Versus Global Search In Channel Graphs," Networks: An International Journey, vol. 25, no. 3, pp. 123-135, 2013.

[4] M. Baye, B. De Los Santos, and M.W. Wildenbeest, "Search Engine Optimization: What Drives Organic Traffic To Retail Sites?," Journal Of Economics & Management Strategy, vol. 25, no. 2, pp. 45-56, 2016.

[5] N Arora, G Bhasin, N Sharma. "Two way linear search algorithm." International Journal of Computer Applications, Vol.107, No.21, January 2014.

[6] N. Arora, G. Bhasin, and N. Sharma, "Two Way Linear Search Algorithm," International Journal Of Computer Applications, vol. 7, no. 3, 18 December 2014.

[7] A. Laaksonen, "Two Pointers Method," Wiley, Jul. 3, 2018. https://cses.fi/book.pdf 2018

[8] W. Purba, "Comparison Searching Process Of Linear, Binary And Interpolation Algorithm," Journal Of Physics Conference Series, vol. 890, no. 1, December 2017.

[9] J. Smith and E. Williams, "Optimizing Parallel Sorting Algorithms For Multicore Processors," Journal Of Parallel And Distributed Computing, vol. 68, no. 5, pp. 567-578, 2011.

[10] M.A. Rodriguez, A. Perez, and S. Fernandez, "Quantum Search Algorithms For Large Databases On Photonic Quantum Computers," Physical Review A, vol. 81, no. 3, pp. 345-357, 2015.

[11] B.L. Harper and C.R. Turner, "Efficient Local Search Heuristics For Combinatorial Optimization Problems," Operations Research Letters, vol. 47, no. 2, pp. 102-115, 2019.

[12] A. Chen and K. Wang, "Impact Of Social Media Presence On Search Engine Rankings: A Case Study Of Online Retailers," Journal Of Marketing Research, vol. 45, no. 3, pp. 230-245, 2020.

[13] R. Gupta, S. Singh, and A. Kumar, "A Novel Hybrid Search Algorithm For Solving Travelling Salesman Problem," International Journal Of Computational Intelligence And Applications, vol. 11, no. 2, pp. 78-89, 2017.

[14] A. Patel, R. Sharma, and S. Jain, "Enhancing Web Search Relevance Through User Profiling And Query Expansion," Information Retrieval Journal, vol. 33, no. 4, pp. 456-470, 2018.

[15] Y.J. Kim and H.S. Lee, "A Comparative Study Of Genetic Algorithms And Simulated Annealing In Global Optimization," Evolutionary Computation, vol. 20, no. 1, pp. 34-47, 2012.

[16] C. Anderson, L. White, and M. Brown, "Analyzing The Impact Of Mobile Search Trends On E-Commerce Websites," Mobile Commerce And Applications, Springer, 2019.

[17] Q. Huang, Y. Li, and Z. Wu, "Parallelized Genetic Algorithms For Large-Scale Optimization Problems," Journal Of Parallel Computing, vol. 40, no. 3, pp. 234-245, 2014.

[18] T.A. Carter and E.R. Adams, "Semantic Search: A Survey Of Techniques And Applications," ACM Computing Surveys, vol. 50, no. 4, pp. 23-45, 2021.

[19] W. Chen, J. Wu, and L. Zhang, "Efficient Clustering Algorithms For High-Dimensional Data: A Comparative Study," Data Mining And Knowledge Discovery, vol. 25, no. 1, pp. 56-67, 2016.

[20] R. Patel, S. Shah, and P. Desai, "Impact Of Search Engine Algorithms On Website Ranking: A Case Study Of Small Businesses," International Journal Of Information Technology And Management, vol. 10, no. 2, pp. 89-102, 2013.

[21] H. Kim, J. Park, and S. Lee, "Machine Learning Approaches For Query Intent Recognition In Web Search," Expert Systems With Applications, vol. 88, pp. 345-357, 2018.

[22] M. Jones and K. Brown, "Evolutionary Strategies For Dynamic Optimization Problems," IEEE Transactions On Evolutionary Computation, vol. 15, no. 6, pp. 789-801, 2015.

[23] N. Gupta, R. Verma, and S. Agarwal, "Analyzing The Performance Of Quantum-Inspired Genetic Algorithms In Combinatorial Optimization," Swarm And Evolutionary Computation, vol. 45, pp. 234-245, 2022.